\documentclass[aps,prl,reprint,amsmath,amssymb,superscriptaddress]{revtex4-1}

\usepackage{graphicx}
\usepackage{bm}

\begin{document}


\title{Stress- and temperature-dependent hysteresis in shear modulus of solid helium}



\author{D. Y. Kim}
\thanks{Both authors contributed equally to this work.}
\affiliation{Center for Supersolid and Quantum Matter Research and Department of Physics, KAIST, Daejeon 305-701, Republic of Korea}

\author{E. S. H. Kang}
\thanks{Both authors contributed equally to this work.}
\affiliation{Center for Supersolid and Quantum Matter Research and Department of Physics, KAIST, Daejeon 305-701, Republic of Korea}

\author{E. Kim}
\affiliation{Center for Supersolid and Quantum Matter Research and Department of Physics, KAIST, Daejeon 305-701, Republic of Korea}

\author{H. C. Kim}
\affiliation{National Fusion Research Institute (NFRI), Daejeon 305-333, Republic of Korea}


\date{\today}

\begin{abstract}
The shear modulus of solid $^{4}$He below 200 mK exhibits an unusual increase, the characteristics of which show remarkable similarities to those of the period reduction in torsional oscillator experiments. We systematically studied the drive strain and temperature dependence of the shear modulus at low temperatures. The hysteretic behavior depends strongly on the drive and cooling history, which can be associated with the thermally assisted Granato--Lucke dislocation theory. The phase diagram of the shear modulus is constructed on the basis of the emerging hysteretic behavior.
\end{abstract}


\maketitle

An anomalous increase in the shear modulus at low temperatures was recently observed in solid helium \cite{day2007}. The anomaly exhibits temperature, frequency, and $^{3}$He concentration dependences resembling those of the non-classical rotational inertia (NCRI) in torsional oscillator (TO) experiments \cite{kc2004na, kc2004sc}. These remarkable similarities have attracted theoretical and experimental attention \cite{chan2008, balibar2010}. For instance, a dislocation vibration model for the NCRI is proposed. The temperature dependence of the period of a torsional oscillator containing solid helium is derived from the variation in the average pinning length of dislocations in solid helium \cite{iwasa2010}. Other efforts to provide a non-supersolid explanation for the TO response were also motivated by the similarities between the two phenomena \cite{chan2008, balibar2010, galli2008, boninsegni2012}. The Cornell group recently reported that the TO motion was controlled by the same microscopic excitation that is agitated by temperature and stress independently, which they suggested can be understood within the framework of a dissipation model in which the elastic properties change at low temperatures \cite{pratt2011}. Nevertheless, the resonant period reduction found in some earlier TO measurements, including Ref. \cite{pratt2011}, can be ascribed to the change in the elastic properties of solid helium filling a torsion rod \cite{beamish2012}. Thus, the similar response of TO to the drive and temperature agitation can be understood in terms of the dislocation pinning mechanism.

The shear modulus increase can be interpreted as the pinning of dislocations by $^{3}$He impurities according to the Granato--Lucke (GL) dislocation theory \cite{gl1956}. The original GL theory considers only the unpinning of dislocations by applied stress without a thermal-fluctuation-induced unbinding mechanism: a large stress reduces the number of pinning points by detaching $^{3}$He impurities from dislocations, thus suppressing the shear modulus. Nevertheless, the effects of stress and temperature on the shear modulus are similar, so its temperature dependence can be explained by the variation in the dislocation loop length with progressive lengthening of freely vibrating dislocation segments due to thermal evaporation of $^{3}$He impurities \cite{day2007, iwasa2010, day2009}. Moreover, the softening of the shear modulus is not a phase transition but a crossover of the thermally activated relaxation process from a stiff (pinned) state to a relaxed (unpinned) state.

The discrepancy in the effects of the drive stress and temperature appears when the hysteretic behaviors during stress and temperature scans are considered. Strong hysteresis is one of the most remarkable characteristics found in both shear modulus and TO response measurements. When a solid sample was cooled at a high TO amplitude, the measured NCRI was small. The NCRI became large when the TO amplitude was reduced at low temperature. Saturation of the NCRI was then detected below the critical amplitude. However, the NCRI did not increase until a substantially larger TO amplitude was applied at low temperature. A similar hysteresis was also reported in a number of TO experiments \cite{aoki2007, clark2008, kim2010NJP, choi2010Na}. The hysteresis between the drive-up sweep and the drive-down sweep disappeared at temperatures above about 60 mK. The relaxation of the resonant period of TO was extended in the narrow region of the phase diagram where the hysteresis was dominant.

Essentially the same drive-dependent hysteretic behaviors appear in the shear modulus anomaly. The hysteresis in the shear modulus was attributed to the motion of dislocations \cite{day2010}. When a solid helium sample was cooled at high amplitudes, the attachment of $^{3}$He impurities to dislocations was hampered by their rapid motion. When the drive was reduced at low temperatures, the binding of the $^{3}$He impurities pinned the dislocations and enhanced the modulus. Impurity binding reduced the pinning length of dislocations, so larger stresses were necessary for unbinding the $^{3}$He impurities from dislocations. On the other hand, no clear thermal hysteresis appeared in the temperature scan. Although a similar agitation effect of temperature and amplitude (stress) was seen in TO (or shear modulus) measurements \cite{pratt2011}, the role of the drive stress and temperature cannot be equivalent, and the true nature of these agitations is not yet clear. Thus, it is important to investigate the thermal and mechanical hysteresis systematically to understand the subtle difference in the effect of drive stress and temperature agitations on the shear modulus. 

Here, we present systematic studies of stress- and temperature-dependent hysteresis in the shear modulus of solid helium. A pair of piezoelectric shear transducers was employed to measure the shear modulus. The gap between the two piezoelectric transducers was 0.4 mm, as explained in a previous report \cite{kim2011}. The shear modulus is proportional to $I/fV$, where $f$ is the frequency of the driving transducer (fixed at 1000 Hz here), $I$ is the induced current at the detecting transducer, and $V$ is the excitation voltage at the driving transducer. A solid helium sample at a pressure of 39 bar was prepared from commercially available $^{4}$He with a nominal isotopic $^{3}$He impurity of 0.3 ppm. We used the blocked capillary method to grow solid helium that was annealed at 1.95 K for 3 h.

\begin{figure}[t]
\includegraphics[width=8.5cm]{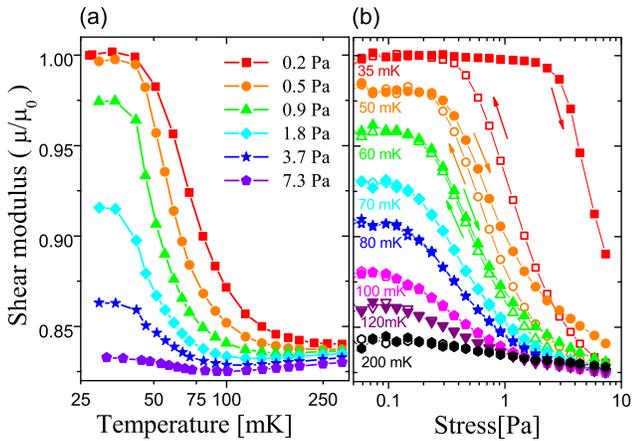}
\caption{\label{fig:CSCT} (a) Shear modulus measured under constant stress scans at various stress values. (b) Shear modulus measured by sweeping stress values down and up at various temperatures.}
\end{figure}

Two different procedures were used to measure the stress- and temperature-dependent hysteresis of the shear modulus. First, the modulus was measured under varying temperature at a constant applied strain (CS). The sample was cooled to 20 mK after undergoing constant strain at 500 mK. The shear modulus was measured during the warming and cooling scans. The warming scan data are only plotted in Fig. \ref{fig:CSCT}(a) since no difference between the two CS scans was found (as also reported by other measurements \cite{kim2010th}); Fig. \ref{fig:CSCT}(a) shows the CS shear modulus change during the temperature scan under various applied strains. The shear modulus exhibited the gradual onset of anomalous stiffening below about 250 mK. Above this temperature, dislocations are unpinned from the impurities because both the thermal energy and strain energy overwhelm the binding energy.

Second, the change in the modulus was measured under a varying applied strain at constant temperature (CT). The strain was set to $3\times10^{-6}$ at 0.5 K, and the solid sample was cooled to various target temperatures. The CT shear modulus was measured as the strain descended stepwise to about $1\times10^{-8}$ (CTdn scan) and ascended subsequently to the original value in discrete steps (CTup scan).

Fig. \ref{fig:CSCT}(b) shows the change in the CT shear modulus during the drive sweep at various temperatures. The values remained almost unchanged during cooling under a high strain of about $10^{-6}$. The subsequent decrease in the strain at low temperatures enhanced the CT modulus, which became saturated at sufficiently low strains. The shear modulus showed two different characteristic behaviors depending on the sample temperature when the strain was subsequently increased. The CT shear modulus obtained with increasing strain (CTup) did not trace that with decreasing strain (CTdn) below 70 mK, whereas the modulus was reproducible above 70 mK. The hysteresis was more prominent at lower temperatures, and the CTup shear modulus was not suppressed from the low-temperature saturated value under a stress of almost 10 Pa.

\begin{figure}[b]
\includegraphics[width=9.5cm]{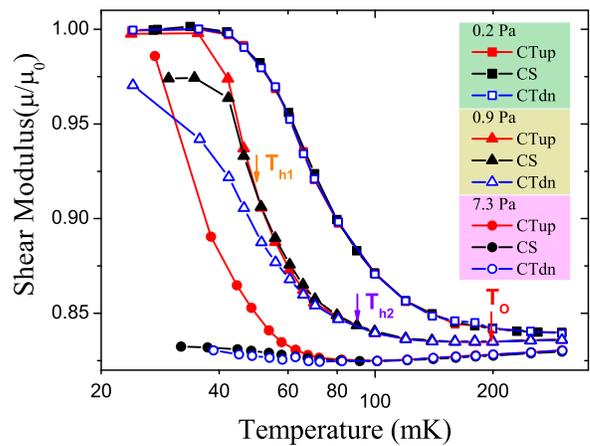}
\caption{\label{fig:SMTemp} Shear modulus with various thermal and mechanical trajectories as a function of temperature. Color codes represent different protocols (thermodynamic histories).}
\end{figure}

Fig. \ref{fig:SMTemp} shows the reconstructed shear moduli of various stresses measured during three thermodynamic trajectories: CS, CTup, and CTdn. The shear moduli measured with stresses smaller than 0.2 Pa collapse onto a single curve. However, the three shear moduli obtained with stresses larger than 0.2 Pa do not agree at low temperatures, although the final stress and temperature are the same.

Above the onset temperature, $T_{o}$, dislocations are completely unpinned. The onset of the pinning process is exceedingly gradual, so it is difficult to determine $T_{o}$. We denoted $T_{o}$ as the temperature at which the shear modulus deviates from the high-temperature linear dependence. $T_{o}$ shows a monotonic drive stress dependence; a larger applied stress interrupts the binding of impurities on dislocations more violently and consequently reduces the onset temperature.

Below $T_{o}$, dislocations become partially pinned. The shear modulus in this partially pinned region is intermediate between the low-temperature saturation value of a strongly pinned (un-relaxed) state and the high-temperature value of a relaxed state. In addition, the shear moduli under different procedures coincide at high temperatures, whereas a history-dependent discrepancy appears at low temperatures. Two more characteristic temperatures related to the inherent discrepancy were identified. As shown in Fig. \ref{fig:SMTemp}, the magnitude of the CS modulus always lies between that of the CTup and CTdn moduli. Above $T_{h2}$, all the thermodynamic trajectories give the same shear modulus. When the sample was cooled below $T_{h2}$, the CTdn modulus first separated from the other two moduli. The CS and CTup moduli did not diverge until the sample temperature was below $T_{h1}$. Namely, all three shear moduli measured under various thermal trajectories show different values below $T_{h1}$.

\begin{figure}[t]
\includegraphics[width=8cm]{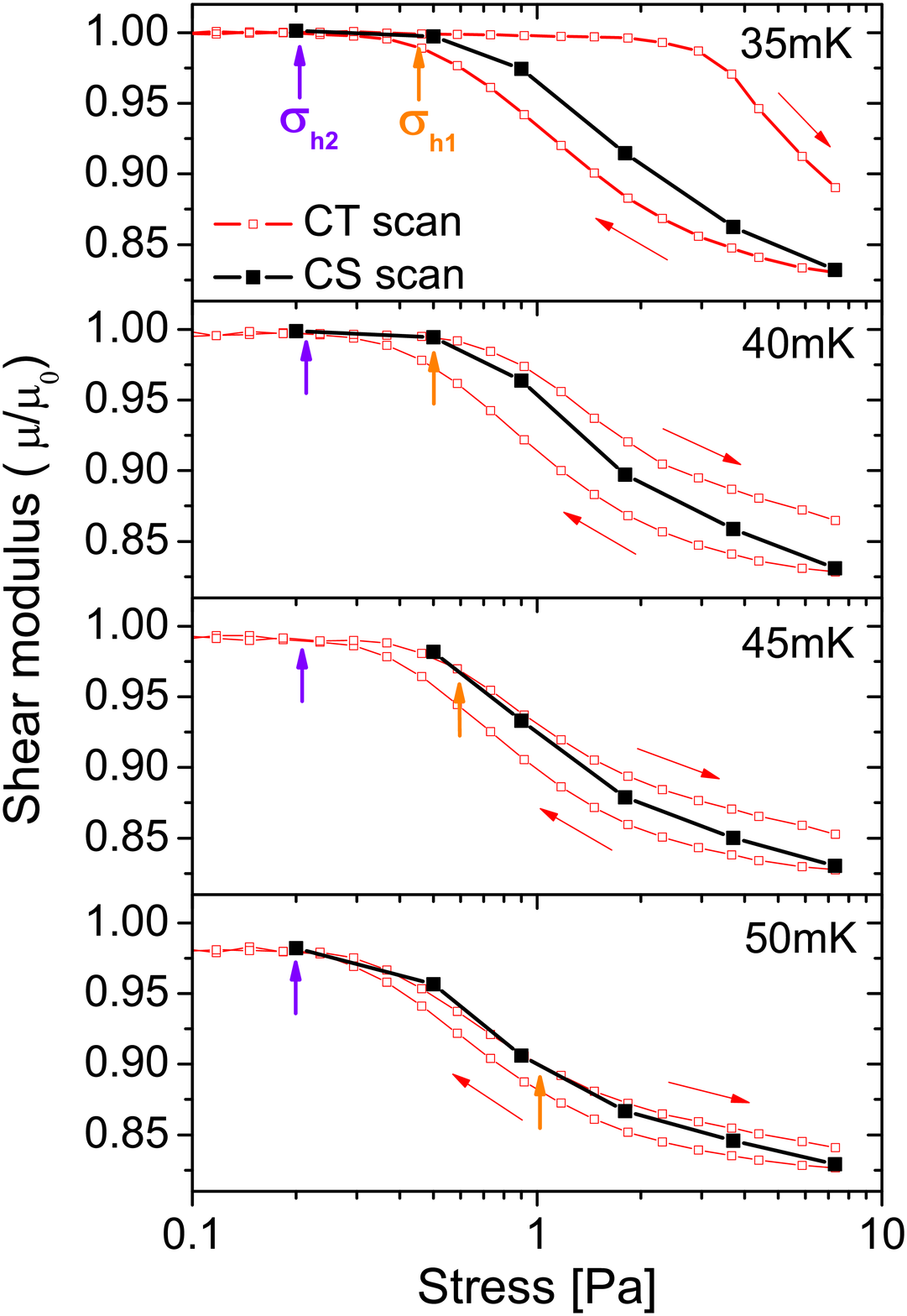}
\caption{\label{fig:SMStress} Shear modulus in solid $^{4}He$ at various temperatures. Red squares are values under decreasing stress (CTdn) and subsequent increasing stress (CTup). Black squares were obtained during CS scan.}
\end{figure}

The discrepancy among the shear moduli with various histories is also pronounced when they are plotted in the shear modulus--stress plane, as shown in Fig. \ref{fig:SMStress}. All the shear moduli with different histories are essentially identical below $\sigma_{h2}$, which is a characteristic stress value measured at a constant temperature in the stress scan and corresponding to $T_{h2}$ measured at the same stress in the temperature scan. Disagreement emerges when the applied stress increases. Between $\sigma_{h2}$ and $\sigma_{h1}$, the CTdn shear modulus deviates from the CTup and CS moduli, which agree. Finally, the CS modulus deviates from the CTup modulus above $\sigma_{h1}$, and all three moduli have different values. The CS and CTdn shear moduli become identical again at a high applied stress, which eventually detaches impurities from dislocations and suppresses the CS modulus. Even under high stresses, the CTup shear modulus remains different from the two other moduli at sufficiently low temperatures.

\begin{figure}[b]
\includegraphics[width=9cm]{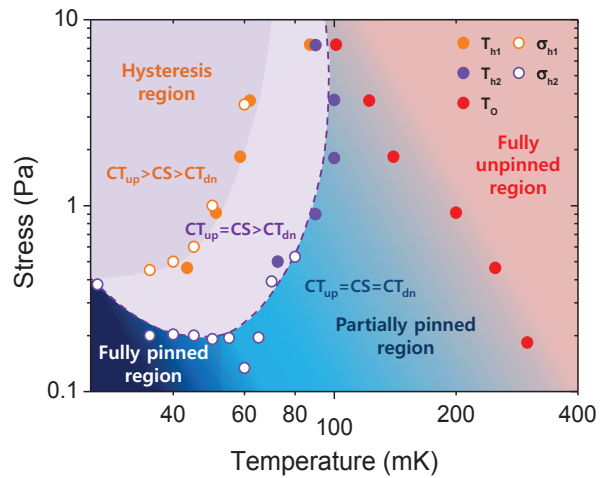}
\caption{\label{fig:diagram} Stress--temperature hysteresis map of solid $^{4}He$ obtained at characteristic points as described in the text.}
\end{figure}

To better understand the effect of stress and temperature, we extracted the characteristic temperatures/stresses at given stresses/temperatures and plotted the hysteresis map in the temperature--stress plane, as shown in Fig. \ref{fig:diagram}. We subtracted the CS modulus from the two CT curves and extrapolated the difference to zero to extract three characteristic temperatures. The set of characteristic points separates the phase diagram into two primary regions: the fully unpinned and partially pinned regions. The latter is further divided into hysteretic and non-hysteretic regions, where the subtle difference between stress and temperature effects is pronounced. Below the boundary, the distinction between the CT and CS shear moduli disappears, whereas the CT modulus differs from the CS modulus above the boundary. The fully pinned state is not available until dislocations are strongly pinned by $^{3}$He impurities at low temperatures and low stresses.

One may try to explain the hysteresis using the dislocation vibration model \cite{iwasa2010}. The hysteresis can be associated with the shortening or lengthening of vibrating segments between two stable nodes of dislocations by $^{3}$He impurities. During the CTdn scan, impurities are bound to dislocations when the dislocation loop length is shorter than the critical length $L_{c}  (\sigma)$. When the scan is reversed to a CTup scan, the previously shortened dislocation loops require greater stress to separate the impurities, which leads to the hysteresis. Thus, the hysteresis appears with variations in the dislocation loop length, which depend strongly on the thermal and mechanical trajectories of solid helium. However, this vibration model is constructed on the naïve assumption that the spacing between $^{3}$He impurities is identical, which cannot explain the details of the hysteresis. For example, once $^{3}$He impurities become unpinned from dislocations, stresses far below the critical value are sufficient to detach neighboring $^{3}$He impurities on dislocations. Accordingly, the breakaway of $^{3}$He impurities may induce a sudden decrease in the shear modulus when the stress exceeds the critical value. One may use the distribution of the impurity pinning length as well as that of the network pinning length to explain the gradual decrease in the CT shear modulus during the drive up-scan \cite{kang2012}.

A significant feature of Fig. \ref{fig:SMTemp} is the discrepancy between the CS and CTdn moduli at low temperatures. The discrepancy is not expected when the stress and temperature have equivalent effects on the shear modulus via the same microscopic excitation. When the stress or temperature decreases, the dislocations are progressively pinned by $^{3}$He impurities. The pinning and unpinning of impurities on dislocations are simply determined by the competition between the pinning potential and the thermal or mechanical energy. Thus, if the thermal and mechanical energy controlled the shear modulus independently, the dependence of the shear modulus on the temperature and the square-root of the strain would show similar responses without a discrepancy \cite{pratt2011}.

Fig. \ref{fig:diagram} provides additional evidence that the stress and temperature affect the pinning of dislocations in different ways. The boundary between the hysteretic and no-hysteresis regions has a positive slope, indicating that a higher temperature is required at higher stress to completely eliminate the hysteresis. This positive slope cannot be understood as well if the effects of temperature and stress are equivalent. If the shear modulus is controlled by a simple combination of thermal and mechanical agitation, the boundary would have a negative slope. Consequently, a complicated interplay of thermal and mechanical influences controls the hysteresis map, and it is crucial to investigate the roles of stress and temperature in order to understand the hysteresis more thoroughly.

We used the thermally assisted breakaway model to describe better both the temperature and drive dependences of the shear modulus change in solid helium at low temperature \cite{teutonico1964, lucke1968}. Assuming that the interaction between an impurity atom and a dislocation is given as a simplified Cottrell force \cite{cottrell1965}, the potential energy of the system consists of three terms: a dislocation tensile line energy and an impurity--dislocation binding energy that likely lead to a pinned state, and the strain energy, which favors an unpinned state \cite{teutonico1964}. The total potential energy of the system has a double-well shape with two minima as a function of the distance between the impurity and the dislocation line: a pinned state at a short distance and an unpinned state at a longer distance. Thermal energy can assist the transition between the two minimum states. The shape of the potential energy is remarkably susceptible to the applied stress \cite{lucke1968}. At low stresses, only one energy minimum appears as a strongly pinned state, and at high stresses, only the unpinned state is energetically stable, indicating a fully unpinned state. In the intermediate range of stresses, two energy minima appear; both pinned and unpinned states are possible.

This intriguing model very clearly explains the hysteresis of the CT shear modulus shown in Fig \ref{fig:CSCT}. If solid helium is cooled under a high applied strain, the shear modulus remains small because the unpinned state is stable under a high stress. As the stress decreases, the second minimum (unpinned state) diminishes, and the system shifts progressively to the pinned state. When the stress is then increased at low temperatures, the second minimum is restored. However, the system cannot transition to an unpinned state if the thermal energy is insufficient to overcome the potential barrier between the two states. This irreversibility produces the discrepancies between the CTdn and CTup moduli at low temperatures. Namely, the hysteresis in the CT modulus can be associated with the stress-dependent effective potential barrier between the two minima. At high temperatures, the absence of the hysteresis can be explained by the presence of sufficient thermal energy, which allows the system to transfer to an unpinned state.

This framework can explain why stress and temperature have different effects. The applied stress changes the shape of the interaction potential; thus, the preferred direction of the process reverses with the sweeping direction. In contrast, the thermal effect enhances the transition probability between the two states without a directional preference. The positive slopes of $T_{h1}$ and $T_{h2}$ as a function of applied stress in the phase diagram reflect the different roles of temperature and stress. A high applied stress enhances the imbalance between the two states in the double-well potential, which causes the difference in the transition probability: the unpinning transition is favored over the repinning transition because of the raised potential barrier from an unpinned to a pinned state. In this partially pinned state, the initial population in the two states and the thermomechanical trajectory that modify the population strongly influence the final state. Thermal fluctuation is the essential ingredient that neutralizes the imbalance. Therefore, higher thermal energy is necessary for compensating the raised potential barrier in the repinning transition.

The subtle difference between the roles of stress and temperature is more pronounced when the CS modulus is compared with CT modulus. Above $T_{h2}$, the sufficiently larger thermal energy compared to the transition barriers allows an unobstructed transition between the two states. Thus, the shear modulus is independent of the stress and temperature history. However, the population in the pinning region during a CTup or CS scan is greater than that during the CTdn scan between $T_{h1}$ and $T_{h2}$ because most dislocations remain in the unpinned state during the CTdn trajectory. The increase in the drive stress deepens the unpinned state in the double-well potential, so the repinning transition from the unpinned state to the pinned state is essentially restricted below $T_{h2}$ owing to the lack of thermal energy. Moreover, the population cannot be augmented simply by lowering the stress in certain temperature ranges. This restraint causes the strongly history-dependent behavior of the modulus.

In contrast, if the passage is chosen to follow the CTup procedure, most dislocations stay in the pinned states. The lack of a thermally assisted crossover significantly suppresses the transition probability to the unpinned state. This observation explains why the CTup modulus always exceeds the CS modulus below $T_{h1}$.

The CS shear modulus manifests the change in the populations of the two states as the thermal fluctuation decreases and is intermediate between the two CT moduli. Above $T_{o}$, the modulus remains at the value of the unpinned state until the system’s thermal fluctuation is reduced sufficiently below the typical binding energy of $^{3}$He impurities and dislocations.

The shape of the shear modulus phase diagram is similar to that of NCRI \cite{choi2010Na}. It is probably reasonable to say that the phenomena are connected, considering their similarities in many properties, including the phase diagram resemblance. One plausible explanation is that the fluctuating mass currents caused by mobile dislocations induce phase fluctuations and destroy the quantum coherence for supersolidity. Thus, the pinning of dislocation lines may be necessary for the appearance of NCRI. All the peculiar properties in the low-temperature anomaly of solid helium also stem from this association. However, the stress values in the NCRI phase diagram are much smaller than those of the shear modulus \cite{kim2011}. Furthermore, an experiment in which DC rotation was superposed on the shear modulus and TO measurements demonstrated a clear contrast between the two phenomena, suggesting that they cannot have the same microscopic origin \cite{choi2010Sc}. Therefore, the underlying relationship between the shear modulus and the TO anomaly is very intricate, and a direct probe of both anomalies, such as simultaneous measurements of both phenomena, is crucial.

\begin{acknowledgments}
We acknowledge support from the National Research Foundation of Korea through the Creative Research Initiatives. 
\end{acknowledgments}

\bibliography{StressT}

\end{document}